\documentclass[preprint,12pt,authoryear]{elsarticle}

\usepackage[colorlinks, citecolor=blue, linkcolor=blue, urlcolor=blue]{hyperref}

\usepackage{tikz}

\usepackage{capt-of, dsfont, nicefrac}
\usepackage{amsmath, amssymb, amsfonts, amsthm, enumerate}
\usepackage{subfig}
\usepackage[normalem]{ulem}

\definecolor{brightpink}{rgb}{1, 0.0, 0.9}
\newcommand{\ignore}[1]{}
\newcommand{\hide}[1]{}
\newcommand{\tus}[1]{{\textcolor{brightpink}{#1 }}{}}
\newcommand{\red}[1]{{\textcolor{red}{#1
}}{}}
\newcommand{\rev}[1]{{\textcolor{black}{#1}}{}}

\newcommand{\Arg}{\mbox{Arg}}
\newcommand{\U}{{\mathcal U}}
\newcommand{\N}{{\mathcal N}}
\newcommand{\A}{{\mathcal A}}
\newcommand{\B}{{\mathbb B}}
\newcommand{\D}{{\mathcal D}}
\newcommand{\act}{{\bf a}}
\newcommand{\E}{{\mathcal E}}
\newcommand{\F}{{\mathcal F}}
\newcommand{\BR}{\text{BR}}

\newtheorem{thm}{Theorem}

\newtheorem{defn}{Definition}
\newtheorem{example}{Example}
\newtheorem{remark}{Remark}
\newtheorem{lemma}{Lemma}
\newtheorem{coro}{Corollary}
\DeclareMathOperator*{\Argmax}{Arg\,max}

\DeclareMathOperator*{\argmax}{arg\,max}

\newtheorem{proposn}{Proposition}

\journal{Games and Economic Behavior}
\begin{document}

\begin{frontmatter}

\title{Equilibrium Cycle: A ``Dynamic'' Equilibrium}

\author[aut_label_tus]{Tushar Shankar Walunj} \ead{tusharwalunj@iitb.ac.in}
\author[aut_label_shik]{Shiksha Singhal} \ead{shiksha.singhal@iitb.ac.in}
\author[aut_label_kav]{Veeraruna Kavitha} \ead{vkavitha@iitb.ac.in}
\author[aut_label_jkn]{Jayakrishnan Nair} \ead{jayakrishnan.nair@ee.iitb.ac.in}
\affiliation[aut_label_tus, aut_label_shik, aut_label_kav]{organization={Industrial Engineering and Operations Research, IIT Bombay},
            addressline={Powai}, 
            city={Mumbai},
            postcode={400076}, 
            state={Maharashtra},
            country={India}}
\affiliation[aut_label_jkn]{organization={Electrical Engineering, IIT Bombay},
            addressline={Powai}, 
            city={Mumbai},
            postcode={400076}, 
            state={Maharashtra},
            country={India}}     
%
\begin{abstract}
In this paper, we introduce a novel equilibrium concept, called the equilibrium cycle, which seeks to capture the outcome of oscillatory game dynamics. Unlike the (pure) Nash equilibrium, which defines a fixed point of mutual best responses, an equilibrium cycle is a set-valued solution concept that can be demonstrated even in games where best responses do not exist (for example, in discontinuous games). The equilibrium cycle identifies a Cartesian product set of action profiles that satisfies three important properties: \textit{stability} against external deviations, \textit{instability} against internal deviations, and \textit{minimality}. This set-valued equilibrium concept generalizes the classical notion of the minimal curb set to discontinuous games. In finite games, the equilibrium cycle is related to strongly connected sink components of the best response graph.
\end{abstract}

\end{frontmatter}

\section{Introduction}

Game theory, as a mathematical framework, helps us to understand the behaviour of rational agents in strategic interactions. A fundamental concept \rev{initially proposed} in game theory is the Nash equilibrium (NE), which was considered as the \emph{outcome} of the game, or indeed, ``the meaning of the game'' \citep{nash1950equilibrium,chain_reccurent_sets2019}. Formally, a Nash equilibrium represents an action profile satisfying the property that no player has a unilateral incentive to deviate from it. 

However, there are well known issues with applying the Nash equilibrium to understand \emph{game dynamics}. By game dynamics, we mean settings where players change their actions over time, seeking to improve their own payoff, in response to the actions of other players. \rev{Such game dynamics, which includes the special case of best response dynamics, do not always converge to a Nash equilibrium~\citep{demichelis2003evolutionary,chain_reccurent_sets2019, benaim2012perturbations}.} From a dynamical systems standpoint, this is not surprising. After all, NEs are simply stationary points (a.k.a., equilibria) corresponding to a broad class of game dynamics, and dynamical systems do not in general converge to a stationary point. Indeed, it is not unusual for game dynamics to result in \emph{limit cycles} (i.e., the action profile approaches a deterministic periodic cycle asymptotically), or even more chaotic behavior in higher dimensions; see~\cite{benaim1999mixed, Ficici_2005_evolutions, benaim2012perturbations, papadimitriou2019MCC,chain_reccurent_sets2019}. In such situations, the Nash equilibrium as a solution concept does not meaningfully capture the `outcome' of the (dynamic) strategic interaction between the players. 

In this paper, we introduce a novel solution concept, which we call the \emph{equilibrium cycle} (EC), that seeks to capture the outcome of oscillatory game dynamics. Specifically, we restrict attention to pure (i.e., non-randomised) player strategies. 
The EC seeks to capture the limit set associated with a broad class of such (oscillatory) game dynamics. Crucially, the definition of the EC does not require the existence of best responses, and is therefore also applicable to discontinuous games, which arise naturally in various contexts \citep{dasgupta1986existence, dasgupta1986existence2, reny1999existence}.\footnote{We first discovered the equilibrium cycle while analyzing a certain pricing game between competing ride-hailing platforms~\citep{uber_paper}. This papers seeks to formalize this equilibrium notion in general terms, provide examples of equilibrium cycles in economics applications, and make connections with other classical equilibrium notions.}

We demonstrate ECs in several well studied games in the economics literature. We motivate the definition of the EC (which is stated formally in Section~\ref{sec:EC_defn}) with the following timing/visibility game from \cite{timing_game_original, timing_game_new}. 

\begin{example}[Visibility game]  \label{ex_timing_game}
Consider a two-player strategic form game, where $\N = \{1,2\}$ represents the set of players, and each player~$i$ has action space $\A_i = [0,1].$ The utility function of player $i$ is defined as
\begin{equation} \label{eqn_timing_game_utility}
\U_i(a_i, a_{-i}) := \begin{cases}
  a_{-i} - a_i, & \text{if } a_i < a_{-i}, \\
  0, & \text{ if } a_i = a_{-i} ,\\
  1 - a_i, & \text{ otherwise.}
\end{cases}    
\end{equation}
\end{example}

This game can be interpreted as a visibility game, where two firms having similar products compete for visibility along a unit-length stretch of highway, with each firm choosing a specific location within this stretch to place its advertising banner. The payoff of each firm depends upon the attention garnered by its banner. The firm that places its banner first along the highway stretch captures the attention of passing vehicles from its chosen location up until the point where the second firm's banner is encountered.
Conversely, if a firm places its banner later along the stretch, it captures the attention of passing vehicles from its location until the end of the highway. In the corner case where both firms place their banners at the same location, it is assumed that neither gains any visibility. An analogous timing interpretation of this game can be found in~\cite{timing_game_new}.

It is easy to observe that this game does not have a pure NE (see~\cite{timing_game_new}). Let us now consider natural `near-best response' dynamics on this game. Note that best responses do not always exist in this game owing to discontinuities in the payoff function. By near-best response, we mean an action that yields nearly-supremal utility for a player in response to the opponent's action. 

Without loss of generality, say we begin with Firm~1 playing action~$a_1 = 0.$ In response, Firm~2, seeking to optimise its payoff, plays the positive action~$\epsilon^{(1)} \approx 0;$ this makes the payoff of Firm~2 nearly maximal (specifically,~$1-\epsilon^{(1)}$), but the payoff of Firm~1 becomes nearly zero (specifically,~$\epsilon^{(1)}$). In response, Firm~1's `better response' is to play action~$\epsilon^{(1)} + \epsilon^{(2)},$ where $\epsilon^{(2)} \approx 0,$ causing its payoff to increase to~$1-\epsilon^{(1)} - \epsilon^{(2)},$ and that of Firm~2 to shrink to~$\epsilon^{(2)}$. In this manner, we see that each firm has an incentive to play an action slightly to the right of its opponent, until any firm's action exceeds~0.5. In particular, once any firm's action exceeds 0.5, the best response to this action is to play the action~0. This `resets' the dynamics, resulting in another cycle as described above.

To summarize, we see that better response dynamics \rev{(among pure strategies)} in the visibility game described in Example~\ref{ex_timing_game} do not converge; rather, the action profile oscillates indefinitely in the set~$[0,0.5]^2$. Moreover, this set~$[0,0.5]^2$ of action profiles satisfies the following properties:

$\bullet$ \emph{Stability}, i.e., given an action profile in this set, neither player has a unilateral incentive to deviate to an `outside' action profile.

$\bullet$ \emph{Unrest}, i.e., given an action profile in the set, at least one player has the incentive to deviate unilaterally to a different `inside' action profile.

$\bullet$ \emph{Minimality}, i.e., no strict subset of this set satisfies the preceding two properties.

As we show in Section~\ref{sec:EC_defn}, the set~$[0,0.5]^2$ is an equilibrium cycle corresponding to the game in Example~\ref{ex_timing_game}. Indeed, the EC is characterized via the above three defining properties: \emph{stability}, \emph{unrest}, and \emph{minimiality}. Intuitively, the first property ensures that under game dynamics, once the action profile enters an EC, it remains in the EC. The second property ensures that the action profile oscillates within the EC indefinitely.  The third property ensures that the EC characterization is tight, i.e., `irrelevant' action profiles are not included within the set.

\rev{
Before we proceed with the formal definition of EC and subsequent analysis, we conclude this section with a discussion on the related literature. 
%
 %
\subsection{Related Literature}
The initial strands of literature analyzing  strategic interactions among multiple agents, predominately describe the outcomes          via   \textit{static concepts}, mainly the Nash Equilibrium \citep{nash1950equilibrium}. After the initial quest, substantial work has examined \textit{dynamic strategic interactions} that more closely reflect the real-world applications.
%
} 

\rev{There are strands of literature that capture the scenarios where finitely many agents \textit{continually   learn the strategies  of  other agents}, in a bid to choose better strategies that   provide them `optimal' utilities. For example,  when the agents primarily consider the most recent moves of the  opponents, we have the  well known best-response dynamics  and when they track significant history of the same we have fictitious play (e.g.,  \cite{brown1951iterative,shapley1963some,binmore2001does}). 
Another branch of literature deals with large population games, where   the agents again \textit{learn better strategies}, but with \textit{more inertia} (they evaluate and revise their policies once in a while) and in a \textit{myopic} manner (the decisions are based on current  state of population choices and are    not so far-sighted). For example, replicator dynamics captures the evolution of population-choice measures  in  large population games like matching or congestion games with imitation protocols  (see e.g., \cite{taylor1978ESS,hofbauer1998evolutionary,sandholm2010population}).} 

\rev{
Several such  learning based studies have established 
the existence of cyclical behavior  (Shapley polygons and limit cycles around a Nash equilibrium, see for example, \cite{brown1951iterative,shapley1963some,matsui1992_best_response_dynamics,samuelson1997evolutionary,binmore2001does, boone2019darwin}).  However,  this part of the literature predominately focuses on   mixed strategies for finite-action games; further, the analysis is derived by    assuming the resultant  dynamical system  to be the solution of an ordinary differential equation or a differential inclusion. Even the recent studies in this direction, that consider some sort of randomness, focus on games with finite actions (e.g., \citet{heinrich2023best} study the dependency on  the outcomes of the dynamics based on the order of player-choices).
While   our  study deals with more generic games  --- also includes the applications with large (compact) action spaces and  with potentially discontinuous utility functions (where  the best responses may not even exist).}

\rev{
Our work more closely relates to other intricate static notions, where   potential outcomes are described using some appropriate mathematical equations/objects, without explicitly describing the dynamics.  One such notion namely    minimal curb set  is introduced by~\cite{basu1991curb} and further related studies are available in~\cite{minimal_curb_sets_2005, hurkens1995learning}. The sink strongly connected components (SCCs) of the better response graph  and the evolution of certain Markov chains  over such graphs are utilized  by \cite{papadimitriou2018srp,papadimitriou2019MCC} to describe the outcomes of the games. 
We would provide more detailed comparison of these notions with EC after introducing the relevant concepts and notations in Sections \ref{sec:connection_to_other_notions} and~\ref{sec:conclusion}; for now, we would like to mention that   EC is a far more general notion that encompasses all the above  notions and can also be utilized to study the discontinuous games.}

\rev{
In \cite{chain_reccurent_sets2019} the authors discuss (and establish) the insufficiency of Nash Equilibrium in representing the outcome of  an arbitrary dynamics related to a finite-action finite-player game in terms of mixed-strategies;  and  our EC is a step forward in providing alternate   outcomes of such games, in fact even for games with large (compact) action sets. }

\rev{
Finally, it has been established that several games mirroring various real-world applications   possess equilibrium cycles:    recommendation systems in  \cite{recommendationsystem},   ride-hailing platforms in \cite{uber_paper}, and supply chains in \cite{wadhwa2025price} are shown to posses equilibrium cycles.
}



\section{Equilibrium Cycle: Definition and examples} \label{sec:EC_defn}

In this section, we formally define the equilibrium cycle and provide some examples.
\ignore{
\begin{defn}[Two player]
Consider a strategic form game $G = \left< \N, (\A_i)_{i \in \N}, (\U_i)_{i \in \N}\right>;$ where $\N = \{1,2\},$ $\A_i$ is the nonempty {\color{red} compact metric space} for every~$i$, and the utility function $\U_i : \A_1 \times \A_2 \to \mathbb{R}.$ A set $\E:= \E_1\times \E_2 \subset \A_1 \times \A_2$ is called an equilibrium cycle if it satisfies the following three properties:
\begin{enumerate}
    \item for any $\act_{-i} \in \E_{-i},$ there exists $a_i \in \E_i$ such that 
    $$\U_i(a_i, \act_{-i}) > \U_i ( \tilde a_i, \act_{-i}) \text{ for all } \tilde a_i \notin \E_i ,$$
    \item if $(a_i, \act_{-i}) \in \E,$ then there exists $i \in \N,$ and $a_i' \in \E_i$ such that $\U_i(a_i', \act_{-i}) > \U_i ( a_i, \act_{-i}),$ and further,
    $$\U_i(a_i', \act_{-i}) > \U_i ( \tilde a_i, \act_{-i}) \text{ for all } \tilde a_i \notin \E_i ,$$
    \item no subset of $\E$ satisfies the above two conditions.
\end{enumerate}
\end{defn}
}
Recall that a strategic form game, also known as a normal form game, represents a simultaneous move game in which all players act simultaneously without knowing the actions of the others. Such a game is formally defined using a tuple $G = \left< \N, (\A_i)_{i\in \N}, (\U_i)_{i\in \N}\right>$, where

 $(i)$ $\N = \{1,\cdots,N\}$ is a finite set of players,

 $(ii)$ $\A_i$ is a nonempty set of available actions for every player~$i$, 

 $(iii)$ $\U_i : \A \to \mathbb{R}$ is the utility (payoff) function of player $i$, with $\A := \prod_{i \in \N} \A_i.$
 
 \vspace{1mm}
 
We use the usual notation: $a_i \in \A_i$ represents an action of  player~$i \in  \N$, $\act_{-i}$ represents a strategy profile of actions of all players except player $i$, and $\act = (a_i, \act_{-i}) \in \A$ represents a strategy profile of all players. Let $\A_{-i} = \prod_{j \in \N; j \ne i}\A_j$ denote the Cartesian product of action sets of all players except player $i.$ 

\begin{defn}[Equilibrium Cycle]
Consider a strategic form game $ \big< \N, (\A_i)_{i \in \N}, $   $(\U_i)_{i \in \N}\big>,$ where for each~$i,$ $\A_i$ is a nonempty set in a metric space.

A closed set $\E:= \E_1 \times \E_2 \times \cdots \times \E_N \subset \A$ is called an equilibrium cycle (EC) if it satisfies the following  conditions:
\begin{enumerate}
    \item \textbf{[Stability] } For any player $i$ and opponent action profile $\act_{-i} \in \E_{-i},$ there exists an action $a_i \in \E_i$ such that 
    $$\U_i(a_i, \act_{-i}) > \U_i ( \tilde a_i, \act_{-i}) \quad \text{for all actions } \tilde a_i \in \A_i \setminus \E_i.$$
    \ignore{\color{red}\item \textbf{[Stability] } for any action profile $\act \in \E$ and any player $i$, there exists $a_i' \in \A_i$ (if required, $a_i' = a_i$) such that $\act':= (a_i', \act_{-i}) \in \E$ and such that
    $$
    \U_i(a_i', \act_{-i}) > \U_i ( \tilde a_i, \act_{-i}) \quad \text{for all exterior profiles, i.e., } \forall \ \tilde a_i \in \A_i ,
 \text{ with } (\tilde a_i, \act_{-i}) \notin \E,    $$}
    \item \textbf{[Unrest] } For any action profile $\act \in \E,$ there exists a player $i \in \N,$ and an alternate action $a_i' \in \E_i$ such that $\U_i(a_i', \act_{-i}) > \U_i ( a_i, \act_{-i}),$ and further,
    $$\U_i(a_i', \act_{-i}) > \U_i ( \tilde a_i, \act_{-i}) \quad \text{for all actions } \tilde a_i \in \A_i \setminus \E_i.$$
    \ignore{\color{red}\item \textbf{[Unrest] } for any interior strategy profile $\act \in \E,$ there exists a player $i \in \N,$ and an alternate action $a_i' \in \A_i$ with $(a_i', \act_{-i}) \in \E$, such that $\U_i(a_i', \act_{-i}) > \U_i ( a_i, \act_{-i}),$ and further such that,
    $$
     \U_i(a_i', \act_{-i}) > \ \U_i ( \tilde a_i, \act_{-i}) \quad \text{for all exterior profiles, i.e., } \forall \ \tilde a_i \in \A_i ,
 \text{ with } (\tilde a_i, \act_{-i}) \notin \E,  
 $$}
    \item \textbf{[Minimality] } No closed~$\F = \F_1 \times \F_2 \times \cdots \times \F_N \subsetneq \E$ satisfies the above two conditions.
\end{enumerate}
\end{defn}

The first condition states that for any player~$i$, so long as all its opponents play actions within their respective EC-components, player~$i$ is also incentivised to play an action within its component~$\E_i$ (specifically, there exists an action within~$\E_i$ that strictly outperforms any `exterior' action). Note that this property is analogous to a pure (strict) Nash equilibrium---no player is incentivised to deviate from its EC-component, so long as the opponents play within theirs. In other words, we have stability against unilateral deviations, now for a \emph{set} of action profiles.  



The second condition states that for any action profile in the EC, there exists at least one player who has a (strict) unilateral incentive to deviate to a different action \emph{within} its EC-component; moreover, this deviation outperforms any `exterior' action. In other words, under any action profile in the EC, at least one player is `unrestful.' 

Together, the first two properties differentiate the EC from the existing equilibrium notions---\emph{the EC exhibits both stability as well as instability}. On one hand, there is stability against unilateral deviations \emph{outside} the EC,  while on the other hand, there is instability/unrest \emph{within}. Finally, the last condition states that an EC is minimal, i.e., no strict subset of an EC is an EC.





%


An immediate consequence of the above definition is that an EC~$\E$ cannot satisfy $|\E| = 1$ (i.e., an EC cannot be a singleton); this follows from the `unrest' condition, whereby for any action profile in the EC, at least one player must have a deviating action providing a strict improvement in utility. Another immediate consequence of the `unrest' condition is the following.

\begin{lemma}\label{lem_no_pure_NE_in_EC}
Consider a game $G$ with an equilibrium cycle~$\E$. 
No action profile $\act \in \E$ is a pure Nash equilibrium.

\end{lemma}

\begin{proof}
Consider any action profile $\act \in \E.$ Invoking the second condition in the EC definition, there exists $i \in \N,$ and $a_i' \in \E_i$ such that $\U_i(a_i', \act_{-i}) > \U_i ( a_i, \act_{-i}).$ Thus, $\act$ is not a pure NE.
\end{proof}
%
%

In the remainder of this section, we provide examples of equilibrium cycles in games that arise in economics applications. We begin with the visibility game described in Example~\ref{ex_timing_game}. 
\begin{proposn}  
 \label{lem_timing_game_EC}
For the visibility game discussed in Example~\ref{ex_timing_game}, the set $[0,0.5]^2$ is an equilibrium cycle.
\end{proposn}
\begin{proof}[Proof of Lemma~\ref{lem_timing_game_EC}]
Fix~$a_{-i} \in [0,0.5].$ We now show that there exists~$a_i \in [0,0.5]$ such that for all~$\tilde{a}_{i} \in  (0.5, 1],$
\begin{equation}
\label{eq:eg1_1}
\U_i(a_{i}, a_{-i}) > \U_i(\tilde a_{i}, a_{-i}) = 1 - \tilde a_{i}.
\end{equation}
If $a_{-i} < 0.5,$ then any choice of $a_i \in (a_{-i},0.5),$ which implies $\U_i(a_{i}, a_{-i}) = 1-a_i,$ satisfies~\eqref{eq:eg1_1}. On the other hand, if $a_{-i} = 0.5,$ then $a_i=0,$ which implies $\U_i(a_{i}, a_{-i}) = 0.5,$ satisfies~\eqref{eq:eg1_1}. This proves that~$[0,0.5]^2$ satisfies the first condition for an EC. 


Next, we prove that $[0,0.5]^2$ satisfies the second condition for an EC. Consider any $(a_1, a_2) \in [0,0.5]^2.$
\begin{itemize}
    \item If $a_1 \neq a_{2}$, without loss of generality, assume $a_i > a_{-i}$. In this case, player~$i$ has an incentive to deviate to an action $a_i' \in (a_{-i}, a_i)$, which yields utility $1 - a_i' > 1 - a_i$. Additionally,    \begin{equation}\label{eqn_timing_proof_second_prop}
    \U_i(a_{i}', a_{-i}) = 1 - a_{i}' > 1 - \tilde a_i =  \U_i(\tilde a_{i}, a_{-i}) \quad \forall\ \tilde a_{i} \in  (0.5, 1].
    \end{equation}
    \item If $a_1 = a_{2}$, both players receive zero utility, and thus any player can deviate and achieve non-zero utility. In particular, if $a_{1} = a_2 = 0.5$, the best response for the deviating player, say player~$i$, is to choose $a_i' = 0$. On the other hand, if $a_1 = a_2 < 0.5$, the deviating player, say player~$i$, can choose $a_i' \in (a_i, 0.5]$, obtaining a positive utility. It is easy to see that in both cases, the deviation satisfies \eqref{eqn_timing_proof_second_prop}.
\end{itemize}
This proves that $[0,0.5]^2$ satisfies the second condition for an EC.

To establish the third condition, suppose, for the purpose of obtaining a contradiction, that~$\F \subsetneq \E$ is a non-empty closed Cartesian product set satisfying the first two conditions for an EC. Define the diagonal $\D := \{(x,x): x \in [0, 0.5] \}$. Note that we cannot have $\D \subset \F,$ since this would imply (given that $\F$ is a Cartesian product) that $\F = \E.$ It follows that $\F \cap \D \subsetneq \D.$ Consider the following cases.

\noindent {\bf Case 1:} $\F \cap \D = \emptyset.$ Consider any $(a_1, a_{2}) \in \F,$ such that  $a_1 > a_{2}$ (without loss of generality). Since~$(a_2,a_2) \notin \F$ and~$\F$ is closed, there exists~$\epsilon > 0$ such that the open ball~$B((a_2,a_2), \epsilon)$ is contained within~$\F^c.$ Now, choosing $a_1' = a_2 + \nicefrac{\epsilon}{2},$ note that $(a'_1,a_2) \in \F^c,$ and 
  $$
    \U_1(a_1', a_2)  = 1 - a_1' > \U_1(\tilde a_1, a_{2}) \ \forall\ \tilde{a}_1 \in \F_1.
    $$
In other words, there exists a deviating action for player~1 that strictly dominates any action within~$\F_1$. This contradicts that~$\F$ satisfies the first condition for an EC.

    
\noindent \textbf{Case~2:} $\F \cap \D \neq \emptyset$. Define
    $$
    b := \sup \{x : (x, x) \in \D \setminus \F \}.
    $$ 
    Observe that $0< b \leq 0.5,$ since $\F$ is closed and $\F \cap \D \neq \emptyset$. We now consider the following sub-cases:
    \begin{enumerate}[$(i)$]
    \item $\{x \in [0,b) \colon (x,x) \in \F\} = \emptyset:$
    Here, by the construction of $b,$ it follows that $(0,0) \notin \F$ and $(0.5,0.5) \in \F,$ and thus there exists a player (w.l.o.g., say player $i$) such that $0 \notin \F_i$ This contradicts the first condition for an EC, since the unique best response of any player~$i$ to the opponent's action~$a_{-i} = 0.5 $ is $0 \notin \F_i.$ 

    \item  $\{x \in [0,b) \colon (x,x) \in \F\} \neq \emptyset:$ In this case, define $$  c :=  \sup \{x : (x, x) \in  \F \cap [0,b)^2\}.
    $$
    By the construction of $b$ and $c$, it follows that $c < b,$ $(c,c)$ lies in $\F,$ and $(x,x) \notin \F$ for all $x \in (c,b).$ Now, for any player~$i,$ fix opponent's action~$a_{-i} = c.$ Then for $a_i \in (c,b),$   
    $$
    \U_i(a_i, c) = 1 - a_i > \U_i(\tilde a_i, c) \ \forall \ \tilde a_i \in \F_i.
    $$
    This contradicts that~$\F$ satisfies the first condition for an EC.
    \end{enumerate}
In summary, we have shown (via a contradiction-based argument) that there does not exist a closed Cartesian product~$\F \subsetneq \E$ that satisfies the first two conditions of an EC. We conclude that $[0,0.5]^2$ is an EC for the visibility game in Example~\ref{ex_timing_game}.

\end{proof}
Interestingly, it is known that the visibility game of Example~\ref{ex_timing_game} admits a \emph{mixed} Nash equilibrium supported on $[0, 1-\nicefrac{1}{e}]^2,$ where~$e$ denotes Euler's number; see~\cite{timing_game_new}. It is instructive to note that the support of the mixed NE differs from the EC, i.e., $[0,0.5]^2.$  

Additionally, we note that visibility game 
can be extended to $N$ players; see~\cite{timing_game_new}. The action space of each player remains~$[0,1].$ Defining $L(i) := \{a_j: a_j \geq a_i \text{ and } j \neq i\},$ the utility of player~$i$ is given by
$$
\U_i(a_1, a_2, \cdots, a_n) = \begin{cases}
    \min\left( L(i) \right) - a_i, & \text{if } L(i) \neq \emptyset, \\
    1 - a_i, & \text{else.}
\end{cases}
$$
It can be shown that $[0, \nicefrac{(N-1)}{N}]^N$ is 
an EC for this $N$ player extension. Observe that the action set corresponding to each player in this  EC grows as $N$ increases.

Next, we consider a two-player Bertrand price competition with an operational cost, and demonstrate an EC therein.

\begin{example}[Bertrand duopoly] \label{ex_bertrand_duopoly}
Consider the following variation of the two-player Bertrand duopoly (see~\cite{osborne_game_theory}) where firms additionally incur a fixed operational cost. Formally, the firms have a fixed marginal cost of production~$c > 0$, and a fixed operational cost~$O_c > 0$ for participating in the market. Each player/firm~$i$ has two options: it can either operate (incurring the cost~$O_c$) by setting price $p_i \geq 0$, or choose not to operate, via the `opt-out' action~$p_i = n_o < 0$ (the opt-out action is represented by a negative number for mathematical convenience), incurring no costs. Thus, the action space of each player is modeled as $[0,\infty) \cup \{n_o\}.$ 

The demand function is assumed to be linear for simplicity: 
\begin{equation*}
D(p) = \begin{cases}
        \alpha - p & \text{if } \ 0 \leq p \leq \alpha,
        \\
        0 & \text{else.} 
    \end{cases}
\end{equation*}
Here, $D(p)$ represents the aggregate demand generated at price~$p.$ When both firms operate, the entire demand is met by the firm offering the lower price. Formally, the utility function of firm~$i$ is given by:
\begin{equation} \label{eqn_bertrand_duopoly_utility}
\U_i(p_i, p_{-i})  =    \begin{cases}
        (p_i - c) D(p_i) - O_c & \text{if } \ p_i \neq n_o\ \text{ and }\ p_i < p_{-i},\\
        \frac{1}{2}(p_i - c) D(p_i) - O_c  & \text{if } \ p_i = p_{-i} \neq n_o,\\
        - O_c  & \text{if }\  p_i > p_{-i} \neq n_o,\\
        0 & \text{if } \ p_i = n_o.
            \end{cases}
\end{equation}
Finally, we assume $\alpha > c + 2\sqrt{O_c}$ to ensure the possibility of a positive utility.
\end{example}

We begin by providing intuition for the EC in Example~\ref{ex_bertrand_duopoly}.
Note that in a monopoly setting, i.e., when only one firm operates, it is easy to see that its optimal (payoff maximizing) price is $p^*_m := \nicefrac{(\alpha + c)}{2}$. 
Similarly, from~\eqref{eqn_bertrand_duopoly_utility}, the break-even price  for a monopolistic firm (i.e., the price at which revenue matches cost, resulting in zero payoff) is 
$$
p_b:= p^*_m - \sqrt{\frac{\left(\alpha\ -\ c\right)^{2}}{4}-O_{c}}.
$$ 
Let us now consider the `near-best response' dynamics for the game in Example~\ref{ex_bertrand_duopoly}. Without loss of generality, suppose that Firm~1 plays the monopoly optimal action $p_1 = p^*_m$. In response, Firm~2 finds it beneficial to set a price slightly lower than $p^*_m$ to capture the entire market (see~\eqref{eqn_bertrand_duopoly_utility}). This causes the payoff of Firm~2 to be near-optimal, but makes the payoff of Firm~1 negative (thanks to the operational cost). In response, Firm~1 is similarly incentivised to undercut Firm~2. This sequence of price reductions between the two firms continues until one of the firms hits the break-even price, where its utility becomes zero. 
Once any firm hits the break-even price, the best response for the opponent is to simply not operate, as any further reduction in price would result in a negative utility. However, once this happens, the firm that was operating at break-even is now incentivised to raise its price back to the monopoly optimal price~$p^*_m,$ and the cycle continues. 
This suggests that the action of each firm oscillates within the set $\left ( \left[p^*_m - \sqrt{\nicefrac{(\alpha - c)^2}{4} - O_c} , p^*_m\right] \cup \{n_o\} \right ).$  This is formalized in the following lemma.

\rev{
\begin{proposn}
\label{lemma:BertrandEC}
Consider the Bertrand duopoly  defined in Example~\ref{ex_bertrand_duopoly}. If~$\alpha > c + 2\sqrt{O_c},$ then
$$
  \E := \left ([p_b, p_m^* ] \cup \{n_o\} \right)^2
$$
  is an equilibrium cycle.
\end{proposn}}
It is instructive to note that in the classical Bertrand duopoly, $O_c = 0,$ resulting in the unique Nash equilibrium $(c, c).$ It is the (reasonable) introduction of a positive operational cost (that is not so large that it prohibits firms from operating at all, as ensured by our assumption) that results in the EC characterized in Proposition~\ref{lemma:BertrandEC}. Finally, we note that the linear demand assumption is not essential to the emergence of the EC; an analogous EC can be established under any smooth decreasing demand function (so long as a positive utility is possible). The proof of Proposition~\ref{lemma:BertrandEC} is analogous to that of Proposition~\ref{lem_timing_game_EC},  and is therefore omitted.

The Bertrand duopoly posits an extreme `all or nothing' bifurcation of market demand between the firms. The following example, which generalizes a pricing game between competing ride-hailing platforms in~\cite{uber_paper}, considers a `less extreme' payoff discontinuity.

\begin{figure}[ht]
\centering
\begin{tikzpicture}
    \draw[->, thick] (0,0) -- (5,0);
    \draw[->, thick] (0,0) -- (0,3.8);
    \draw[-, thick] (4.5, -0.1) node[below] {$a_M$} -- (4.5, .1);
    
    \draw[dashed, thick] (1.29586, -0.25)  node[right] {$b$} -- (1.29586,3.6);
    \draw[dashed, thick] (3,-0.25) node[right] {$c$} -- (3,3.6) ;

    \draw[dashed, thick] (1.29586,1.54) -- (3,1.54);

    \draw[red, very thick, domain=0:4.5, smooth] plot (\x, {1.53 - 0.17*(\x - 3)^2});
    \draw[blue, very thick, domain=0:4.5, smooth] plot (\x, {1.5*\x*exp(-0.175*\x)});

    \node[blue] at (4,3.3) {$f$};
    \node[red] at (4,1.6) {$g$};
    \node[rotate=90] at (-0.3, 1.8) {Utility};
    \node[] at (2.5, -0.7) {Price};
    \node[below] at (0,0) {$0$};

\end{tikzpicture}
\caption{Illustrations of the functions $f$ and $g$ in Example~\ref{ex_UBER_simialar}.}
\label{fig:uber_diag}
\vspace{-2mm}
\end{figure}
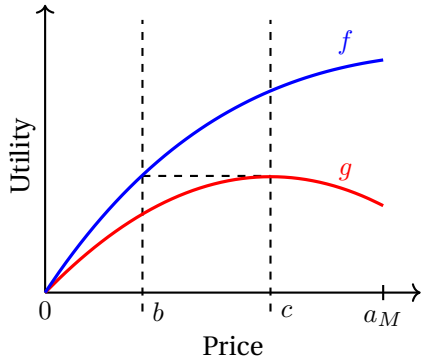

\begin{example}\label{ex_UBER_simialar}
Consider a symmetric two-player game $ G = \left \langle \mathcal{N}, \mathcal{A},\ \mathcal{U} \right \rangle.$ For each player~$i,$  $\mathcal{A}_i = [0, a_{M}],$ where $a_M > 0.$ The utility function for player $i$ is defined as follows.
$$
U_i(a_i, \act_{-i}) = \begin{cases}
    f(a_i) & \text{if } a_i < a_{-i} ,\\
    g(a_i) & \text{if } a_i > a_{-i} ,\\
    \frac{f(a_i) + g(a_i)}{2} & \text{else.}
    \end{cases}
$$

Here, $ f: \mathcal{A}_i \to \mathbb{R} $ is a continuous and strictly increasing function, and $ g: \mathcal{A}_i \to \mathbb{R} $ is a continuous and strictly concave function. 
Additionally, $f(0) = g(0)$ and $f(a) > g(a)$ for all $ a \in (0,a_M];$ see Figure~\ref{fig:uber_diag} for an illustration.
%
%
%

\end{example}

The game in Example~\ref{ex_UBER_simialar} also admits an EC, as shown in the following lemma; this generalizes Theorem~4 in \cite{uber_paper}. The proof is once again omitted, given its similarity to the proof of Proposition~\ref{lem_timing_game_EC}. 
\begin{proposn}
Consider the game described in Example~\ref{ex_UBER_simialar}. Let $b := f^{-1}(g(b))$ and 
$
c:= \argmax_{a_i \in \A_i} g(a_i),$   Then $\E = [b,c]^2$ is an equilibrium cycle of this game.
\end{proposn}

Finally, we consider the following two-player discrete-action game from~\cite{papadimitriou2019MCC}.

\begin{example} \label{ex_discrete_game}
Consider a two-player game defined in Table~\ref{tab:discrete_game_example} (see~\cite{papadimitriou2019MCC}), where player~$1$ has the action set $\{U, M, D\}$ and player~$2$ has the action set $\{L, C, R\}$:

\begin{table}[ht!]
    \centering
        \begin{tabular}{|c|c|c|c|}
            \hline
            & L & C & R  \\ 
            \hline
            U & (2,0) & (0,2) & (0,0) \\ \hline
            M & (0,2) & (2,0) & (0,0) \\ \hline
            D & (0,0) & (0,0) & (1,1) \\
            \hline
        \end{tabular}
    \caption{Discrete game containing both an EC and a pure NE.}
    \label{tab:discrete_game_example}
\end{table}
\end{example}

It is easy to see in this case that $\{U, M\} \times \{L, C\}$ is an equilibrium cycle; indeed, best response dynamics would oscillate indefinitely within this set. Note that the topological considerations that came up in the preceding examples (where the action space was continuous) do not arise in discrete games such as this one; every subset of the action space is closed here. Interestingly, this game also admits the pure Nash equilibrium~$(D,R)$ (which naturally is outside the EC, consistent with Lemma~\ref{lem_no_pure_NE_in_EC}).

\section{Connection with other equilibrium notions} \label{sec:connection_to_other_notions}

The preceding section makes comparisons between the EC and the Nash equilibrium (pure as well as mixed). In this section, we compare the EC with other equilibrium notions in the literature that have a dynamic connotation. In the special class  of best response games, we show that ECs are intimately tied to curb sets~\citep{basu1991curb}. Specializing further to finite games, we show that ECs are closely tied to strongly connected sink components of the best response graph.


\subsection{Curb sets in best response games} \label{subsec:curb_sets}

We begin with the notion of curb sets, introduced in \cite{basu1991curb}. These sets are defined for games where best responses exist (for example, discrete games, and games where the payoff functions are continuous with respect to a compact action space). We refer to this class of games as BR games, defined formally as follows.
%
%

\begin{defn}[Best Response (BR) game]
We call a game~$G$ a BR game, if, for any player~$i$ and any opponents' action profile~$a_{-i},$ the set of best responses of player~$i,$ denoted $\BR_i(\act_{-i}),$ exists.
\end{defn}

A curb set is defined as a set of strategy profiles that is closed under rational behavior, i.e., a set that `contains its best responses' \citep{basu1991curb}.


\begin{defn}[Curb Set] \label{defn_minimal_curb_set}
Consider a game~$G = \left \langle \mathcal{N}, \mathcal{A},\ \mathcal{U} \right \rangle.$ A non-empty Cartesian product $C = \prod_{i=1}^n C_i \subset \prod_{i=1}^n \A_i$ is a curb set corresponding to this game if 
\begin{equation} \label{eqn_curb_set_defn}
\left( \prod_{i=1}^n \BR_i(C_{-i}) \right) \subset C, 
\end{equation}  
where $\BR_i(C_{-i}) := \cup_{\act_{-i} \in C_{-i} }  \BR_i(\act_{-i})$ for $C_{-i} \subset \A_{-i}.$ A curb set is minimal if none of its strict subsets is a curb set. 
\end{defn}
Note that by definition, a curb set may contain a pure Nash equilibrium, or even a continuum of Nash equilibria. To make the connection between curb sets and equilibrium cycles, we need to introduce the notion of \emph{non-trivial curb sets}.

%
\begin{defn}[Non-trivial curb set]
A curb set $C$ corresponding  to a game~$G$ is non-trivial if it does not contain any pure Nash equilibria. 
\end{defn}
Intuitively, the absence of pure NE induces `unrest' within a curb set, enabling the following equivalence.
%
\begin{thm}\label{thm_cont_game_C=EC}
Consider a BR game $G.$
\begin{enumerate}[(i)]
    \item If $C$ is a non-trivial minimal curb set of~$G,$ then $C$ is also an equilibrium cycle of $G.$
    \item If $\E$ is an equilibrium cycle of $G,$ then $\E$ is also a non-trivial minimal curb set of~$G.$ 
\end{enumerate}
\end{thm}

{\begin{proof}
To prove Part~(i), suppose $G$ has a non-trivial minimal curb set~$C$.
\begin{enumerate}
    \item To prove that $C$ satisfies the first property of an EC, consider any player~$i$ and $\act_{-i} \in C_{-i}.$ Using the definition of the curb set (see \eqref{eqn_curb_set_defn}),
    against $\act_{-i},$ there exists a best response $a_i \in C_i$ such that
    $$
     \U_i(a_i, \act_{-i}) > \U_i(\tilde a_i, \act_{-i})  \text{ for all } \tilde a_i \notin C_i
    $$
    (indeed, note that no other outside action  $\tilde a_i \notin C_i$ is a best response). Thus, the first property of EC is satisfied.
    \item Consider any strategy profile $\act \in C$. Since $C$ is a non-trivial curb set, $\act$ is not an NE, and thus there exists at least one player (say player $i$) which stands to obtain a strictly better utility via unilateral deviation. Suppose $a_i'$ is a best response of this player $i;$ and we have $\U_i(a_i', \act_{-i}) > \U(a_i, \act_{-i})$.  Further, by the definition of curb set, $a_i' \in C_i$ and as before,
    $$
     \U_i(a_i', \act_{-i}) > \U_i(\tilde a_i, \act_{-i})  \text{ for all } \tilde a_i \notin C_i.
    $$
    Thus, $C$ satisfies the second property of an EC.
    
    \item We prove that $C$ satisfies the third property of an EC via a contradiction-based argument. Specifically, suppose that there exists $\E \subsetneq C$ satisfying the first two properties of an EC. Consider any player $i$. Against any $\act_{-i} \in \E_{-i}$, from the first property of EC, there exists $a_i' \in \E_i$ such that
    $$
    \U_i(a_i', \act_{-i}) > \U_i ( \tilde a_i, \act_{-i})  \text{ for all } \tilde a_i \notin \E_i .
    $$
    This implies that the best responses of player~$i$ must be contained in $\E_i,$ which in turn implies that the set $\E$ contains all its best responses. This means that $\E \subsetneq C$ is also a curb set, thus contradicting the assumption that~$C$ is a \emph{minimal} curb set. 
\end{enumerate}
This proves that the nontrivial minimal curb set $C$ is an EC.

To prove Part~(ii), suppose $G$ has an EC $\E$. Using an argument similar to that used in Case~(3) of Part~(i) above, we have that $\E$ is a curb set. Further, using Lemma~\ref{lem_no_pure_NE_in_EC}, we have that $\E$ is a \emph{non-trivial} curb set. All that remains now is to show that $\E$ is also minimal. Towards this, suppose there exists a non-trivial curb set~$C$ such that $C \subsetneq \E$. The argument used in Cases~(1) and (2) of Part~(i) above\footnote{Observe that the minimality of the non-trivial curb set is not used in the first two cases of the proof of Part~(i).} imply that $C$ then satisfies the first two properties of an EC. This contradicts the assumption that $\E$ is an EC.

Thus an EC $\E$ is a non-trivial minimal curb set.
%
\end{proof}}

Theorem~\ref{thm_cont_game_C=EC} establishes that in BR games, equilibrium cycles are equivalent to non-trivial minimal curb sets. However, it is important to note that the equilibrium cycle can be defined even in non-BR games (for example, the discontinuous non-BR games considered in Examples~\ref{ex_timing_game}--\ref{ex_UBER_simialar} in Section~\ref{sec:EC_defn}). Thus, the EC may be interpreted as a generalization of curb sets to non-BR games, albeit with the additional imposition of `unrest' (or non-triviality), which (potentially) manifests as oscillations in a dynamical setting.

We conclude this discussion with the following connection between the EC and the mixed NE, which follows directly from the above equivalence and the results of~\cite{basu1991curb} (see Section~3 therein).



\begin{lemma}
Consider a BR game. If this game has an equilibrium cycle~$\E,$ then there exists a mixed Nash equilibrium of this game with support $\F \subset \E$.
\end{lemma}

\subsection{Connected components of the best response graph in finite games}

Having considered best response games in Section~\ref{subsec:curb_sets}, we now further specialize to finite games. In finite games, it is known that best response dynamics can be understood via the \emph{best response graph} (a.k.a., \emph{best reply graph})~\citep{young1993evolution}. In this section, we establish a connection between equilibrium cycles and certain strongly connected components of the best response graph.

We begin with some definitions.



%
%
\begin{defn}[Best response graph]
In a finite game~$G$, the best response graph is a directed graph with node set~$\A,$ i.e., its nodes represent the pure action profiles of the game. A directed edge exists from node $\act$ to node $\act'$ if and only if: (1) $\act$ and $\act'$ differ only in the action of a single player, say player $i$, such that $a_i \neq a_i'$ and $\act_{-i} = \act_{-i}'$, and (2) $a_i'$ is a best response for player~$i$ against $\act_{-i},$ and strictly preferred over~$a_i,$ i.e., 
$$
\U_i(a_i', \act_{-i}) = \max_{\tilde a_i \in \A_i} \U_i(\tilde a_i, \act_{-i})
\ \text{ and }\ \
\U_i(a_i', \act_{-i}) > \U_i(\act).
$$
\end{defn}



\begin{defn}[sink SCC]
A strongly connected component (SCC) of a directed graph is a maximal subgraph in which any two nodes are mutually reachable. Sink SCCs are SCCs that satisfy the property that the graph contains no edge from a node within the SCC to a node outside the SCC.
\end{defn}

It is well known in graph theory that any directed graph decomposes into disjoint SCCs (see~\cite{cormen2022graph}) and that random walks on the graph almost surely end up in a sink SCC. In the context of finite games, it is easy to see that a pure Nash equilibrium is a singleton sink SCC (i.e., a sink SCC consisting of a single node) of the best response graph. On the other hand, a non-singleton sink SCC of the best response graph represents a limit cycle of best response dynamics; this implies the following connection with the equilibrium cycle.



\begin{thm}
\label{thm:EC-sinkSCC}
Consider a finite game~$G$.
\begin{enumerate}[(i)]
    \item If the best response graph of $G$ contains a non-singleton sink SCC having 
    node set $S = S_1 \times S_2 \times \cdots \times S_N \subset \A$, then $S$ is an equilibrium cycle of~$G.$
    \item If $G$ contains an equilibrium cycle~$\E$, then~$\E$ includes the node set of a non-singleton sink SCC of the best response graph of~$G$.
\end{enumerate}
\end{thm}

\begin{proof}
To prove Part~(i), suppose $S = S_1 \times S_2 \times \cdots \times S_N$ denote the node set corresponding to  a non-singleton sink SCC of the best response graph. We now prove that $S$ satisfies the three conditions of an EC.
\begin{itemize}
    \item  Consider any player $i$ and $\act_{-i} \in S_{-i}.$ Since there is no outgoing edge in the best response graph from any node in~$S$ to a node outside~$S,$ it follows that~$\A_i \setminus S_i$ does not contain a best response for player~$i$ against~$\act_{-i}.$ Since~$G$ is finite, this further implies that a best response~$a_i \in S_i$ exists, so that $\U_i(a_i,\act_{-i}) > \U_i(\tilde{a}_i,\act_{-i})$ for all~$\tilde{a}_i \in \A_i \setminus S_i.$ This establishes that~$S$ satisfies the first condition for an EC.

    \item Consider $\act \in S$. Since the sink SCC is non-singleton, there exists a outgoing edge from node~$\act$ to, say node $\act' \in S.$  Note that these nodes differ only in the action of one player, say player $i$. Thus, $\U_i(a_i', \act_{-i}) > \U_i ( a_i, \act_{-i}).$ Moreover, since there does not exist an outgoing edge from~$\act$ to any node outside~$S,$ $\U_i(a_i', \act_{-i}) > \U_i ( \tilde a_i, \act_{-i}) \ \forall \  \tilde a_i \in \A_i \setminus S_i.$ This establishes that~$S$ satisfies the second condition for an EC.
    
    \item To establish the third condition, we use a contradiction based argument. Suppose that there exists a strict Cartesian product subset~$\F \subsetneq S$ satisfying the first two conditions for an EC. Since the sink SCC (with node set $S$) is strongly connected, there exists 
    $\act \in \F$ and $\act' \in S \setminus \F$ be such that there exists a directed edge from $\act$ to $\act'$ in the best response graph. However, this contradicts that~$\F$ satisfies the first condition for an EC. This establishes that~$S$ also satisfies the minimality condition.
    
\end{itemize}

To prove Part~(ii), consider an EC~$\E.$ From the first condition of an EC, it follows that the best response graph contains no edge from a node in~$\E$ to a node outside~$\E.$  This implies that~$\E$ contains (the node set of) a sink SCC. Moreover, since $\E$ does not contain any pure NE (see Lemma~\ref{lem_no_pure_NE_in_EC}), it cannot contain a singleton sink SCC. Thus, $\E$ contains the (node set of) a non-singleton sink SCC. 
\end{proof}

Theorem~\ref{thm:EC-sinkSCC} shows that in a finite game, a `rectangular sink SCC' (formally, a sink SCC with a node set that is a Cartesian product) of the best response graph supports an EC over its node set. On the other hand, an EC of a finite game contains within it the node set of a non-singleton sink SCC. The dichotomy between these two statements stems from the fact that while an EC is `rectangular' by definition, the node set of a sink SCC of the best response graph need not be so. We illustrate this via the following examples.
 
\begin{example}
\label{ex:discrete_visiblity_game}
Consider a discretized version of the two player symmetric visibility game considered in Example~\ref{ex_timing_game}, where the action space of each player is (uniformly) discretized to $\A_i^{(n)} = \{0, \nicefrac{1}{n}, \nicefrac{2}{n}, \cdots, 1\}.$ This game, parameterized by the discretization parameter~$n \geq 2,$ has the same utility function as before, defined in~\eqref{eqn_timing_game_utility}.
\end{example}

\begin{figure}[ht]
    \centering
    \subfloat[\centering $\A_i^{(6)} = \{0, \nicefrac{1}{6}, \nicefrac{2}{6}, \cdots, 1\}$    \label{subfig:best_response_dynamics_n_6}
    ]{
    \includegraphics[page = 3, trim = {6.7cm 3.4cm 8.5cm 1.9cm}, clip, scale = 0.27]{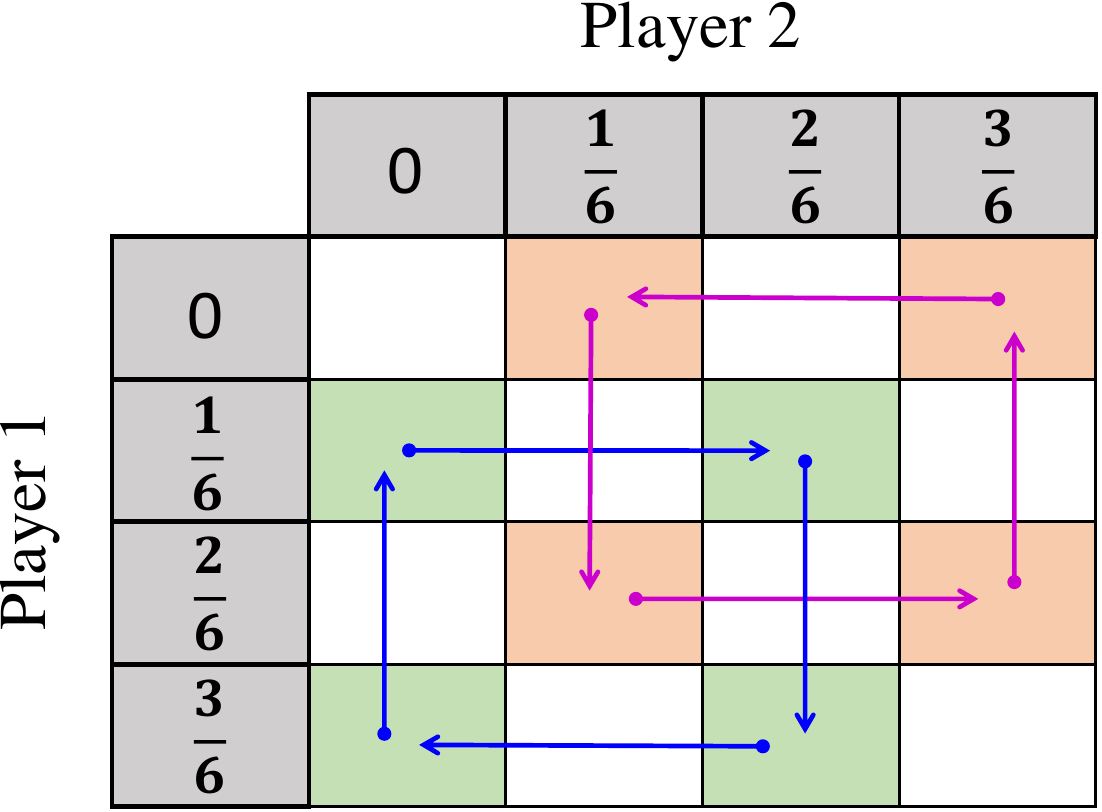}
    }%
    \hspace{.5cm}
    \subfloat[\centering $\A_i^{(7)} = \{0, \nicefrac{1}{7}, \nicefrac{2}{7}, \cdots, 1\}$ \label{subfig:best_response_dynamics_n_7}
    ]{
    \includegraphics[page = 2, trim = {5cm 2cm 6.8cm 1.1cm}, clip, scale = 0.27]{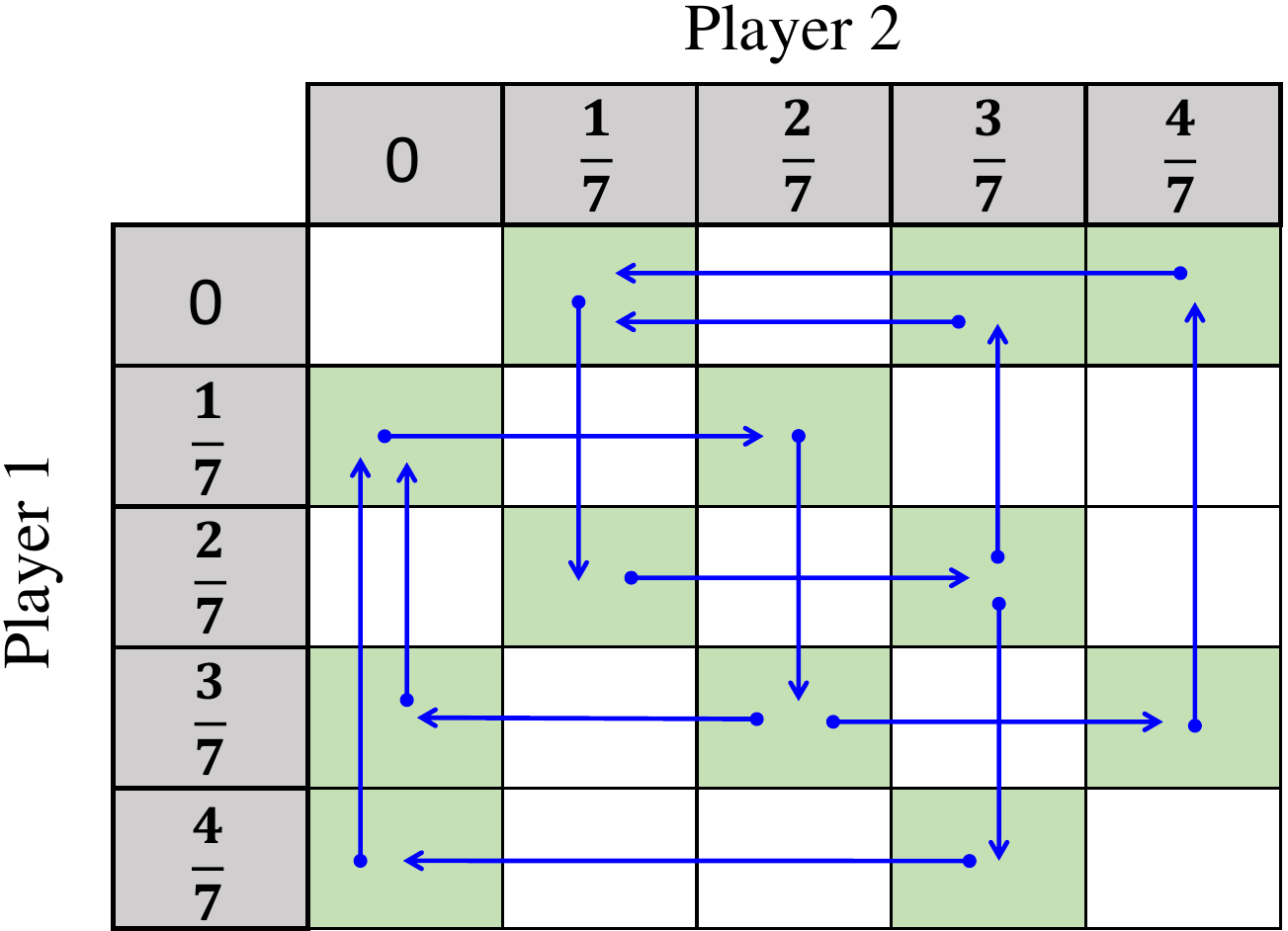}
    }%
    \caption{Sink SCCs for discretized version of visibility game in Example~\ref{ex:discrete_visiblity_game}}
    \label{fig:best_response_dynam ics}
\end{figure}

Let us first consider the above discretized visibility game with~$n = 6.$ In this case, the game admits two ECs, which are both also the node sets of sink SCCs of the best response graph; see Figure~\ref{subfig:best_response_dynamics_n_6}. The green action profiles in the figure represent one EC, and the orange ones represent the other; we also depict the edges of the two sink SCCs in the same figure. Intuitively, best response dynamics would eventually oscillate over either one of these two ECs, depending on how the dynamics are initialized.

\ignore{ 
For discretised version of visibility game with action space $\A_i = \{0, \nicefrac{1}{6}, \nicefrac{2}{6}, \cdots, 1\}$, then we have following two disjoint ECs which are also sink SCCs:
{
$$
\left \{\left(\frac{1}{6}, 0\right), \left(\frac{1}{6}, \frac{2}{6}\right), \left(\frac{3}{6}, \frac{2}{6}\right), \left(\frac{3}{6}, 0\right) \right \}, \text{ and } 
\left \{\left(0, \frac{1}{6}\right), \left(\frac{2}{6}, \frac{1}{6}\right), \left(\frac{2}{6}, \frac{3}{6}\right), \left(0, \frac{3}{6}\right) \right \} .
$$
}
}

It is also instructive to consider the discretized visibility game with~$n = 7.$ This game admits the unique EC $\E = \{0, \nicefrac{1}{7}, \nicefrac{2}{7},  \nicefrac{3}{7},  \nicefrac{4}{7}\}^2$. However, the node set of the unique sink SCC (of the best response graph) is a strict, non-rectangular subset of this EC; the nodes of this sink SCC are marked green in Figure~\ref{subfig:best_response_dynamics_n_7}, with the arrows depicting the edges.

\ignore{ 
For discretised version of visibility game with action space $\A_i = \{0, \nicefrac{1}{7}, \nicefrac{2}{7}, \cdots, 1\}$, then we have the EC set  The sink SCC for this game is as follows:
{
\begin{align*}
S =  \Bigg \{
& \left(\frac{1}{7}, 0\right), \left(\frac{1}{7}, \frac{2}{7}\right), \left(\frac{3}{7}, \frac{2}{7}\right), \left(\frac{3}{7}, \frac{4}{7}\right), \left(0, \frac{4}{7}\right),          \left(0, \frac{1}{7}\right), \\
& \left(\frac{2}{7}, \frac{1}{7}\right), \left(\frac{2}{7}, \frac{3}{7}\right), \left(\frac{4}{7}, \frac{3}{7}\right), \left(\frac{4}{7}, 0\right),     \left(\frac{3}{7}, 0\right),          \left(0, \frac{3}{7}\right) 
\Bigg \}    .
\end{align*}
}
} 

It is also possible to make a connection between ECs and sink SCCs of the \emph{better response graph} \citep{best_response_graph_2023}; this latter object has been studied recently in the context of game dynamics in \cite{papadimitriou2019MCC}.\footnote{The better response graph of a finite game~$G$ is a directed graph over~$\A,$ 
where a directed edge exists between $\act$ and $\act'$ if these two action profiles differ in the action of the single player, say player~$i,$ and player~$i$ strictly benefits from the unilateral deviation to~$\act',$ i.e., $\U_i(a_i', \act_{-i}) > \U_i(\act)$.
} Formally, it can be shown that the node set of a rectangular sink SCC of the better response graph contains an EC. However, a reverse connection does not hold---an EC need not contain the node set of a sink SCC of the better response graph.

\section{Final remarks on the equilibrium cycle}

In this section, we show that ECs of a game do not intersect, and introduce the notion of dominant ECs (analogous to dominant NEs). We begin with the former result.

\begin{thm}\label{thm_ECs_do_not_intersect}
Two equilibrium cycles of a game do not intersect.
\end{thm}
\begin{proof}
Assume, for the purpose of arriving at a contradiction, that there exist two ECs, $\E$ and $\F$ with $\E \cap \F \neq \emptyset.$ The trivial cases $\E \subset \F$ or $\F \subset \E$ violate the third condition (minimality) of the definition of an EC. Therefore, we must have $\E \setminus \F \neq \emptyset$ as well as  $\F \setminus \E \neq \emptyset$. We now prove that $\E \cap \F$ also satisfies the first two properties of an EC; note that this would lead to the desired contradiction (observe that $\E \cap \F$ is also a Cartesian product set). 

To prove that $\E \cap \F$ satisfies the first property of an EC, consider any $i$, and $\act_{-i} \in \E_{-i} \cap \F_{-i}.$ Now, we have to show that against $\act_{-i},$ there exists an action $a_i \in \E_i \cap \F_i$ such that 
\begin{equation} \label{eqn_EC_intersec_five}
    \U_i(a_i, \act_{-i}) > \U_i ( \tilde a_i, \act_{-i}) \quad \text{for all } \tilde a_i \notin  \E_i \cap \F_i.
\end{equation} 
To prove the above claim, consider the following two cases.\\
\textbf{Case 1:} $\E_i \subset \F_i$ or $\F_i \subset \E_i$. This case is trivial; the existence of $a_i$ satisfying \eqref{eqn_EC_intersec_five} follows by invoking the first property of an EC, applied to $\E$ if $\E_i \subset \F_i,$ or to $\F$ if $\F_i \subset \E_i.$\\
\textbf{Case 2:} $\E_i \setminus \F_i \neq \emptyset,$ and $\F_i \setminus \E_i \neq \emptyset$.
    Since $\E$ is an EC, there exists $a_i' \in \E_i $ such that
    \begin{equation} \label{eqn_EC_intersec_two}
        \U_i( a_i', \act_{-i}) > \U_i ( \tilde a_i, \act_{-i}) \quad \text{for all } \tilde a_i \notin  \E_i.
    \end{equation}
    Since $\F$ is also an EC, there exists another $  a_i'' \in \F_i$ such that 
    \begin{equation} \label{eqn_EC_intersec_four}
       \U_i( a_i'', \act_{-i}) > \U_i ( \tilde a_i, \act_{-i}) \quad \text{for all } \tilde a_i \notin  \F_i. 
    \end{equation}
    Now we consider the following four sub-cases:
    \begin{enumerate}[(i)]
        \item $a_i' \in \E_i \setminus \F_i,$ and $a_i'' \in \F_i \setminus \E_i.$ This case is impossible, since \eqref{eqn_EC_intersec_two} implies $\U_i( a_i', \act_{-i}) > \U_i ( a_i'', \act_{-i}),$ whereas \eqref{eqn_EC_intersec_four}, implies $\U_i( a_i'', \act_{-i}) > \U_i ( a_i', \act_{-i}).$
        \item $a_i' \in \E_i \setminus \F_i,$ and $a_i'' \in \F_i \cap \E_i.$ Then using \eqref{eqn_EC_intersec_four}, we have $\U_i( a_i'', \act_{-i}) > \U_i ( a_i', \act_{-i}).$ Thus, by choosing $a_i = a_i'',$~\eqref{eqn_EC_intersec_five} holds.
        \item $a_i' \in \E_i \cap \F_i,$ and $a_i'' \in \F_i \setminus \E_i.$ This case can be handled analogously as above. 
        \item $a_i' \in \E_i \cap \F_i$ and $a_i'' \in \F_i \cap \E_i$. If $\U_i(a_i', \act_{-i}) \geq \U_i(a_i'', \act_{-i}),$ choose~$a_i = a_i',$ and otherwise choose $a_i = a_i''$. It is easy to see that this choice satisfies~\eqref{eqn_EC_intersec_five}.
    \end{enumerate}

To prove that $\E \cap \F$ satisfies the second property of an EC, consider any $\act \in \E \cap \F.$ Now, we have to show that there exists a player~$i,$ and an alternate  action $a_i' \in \E_i \cap \F_i$ such that
\begin{equation} \label{eqn_EC_intersec_six}
    \U_i(a_i', \act_{-i}) > \U_i ( a_i, \act_{-i}), \text{ and } \ \U_i(a_i', \act_{-i}) > \U_i ( \tilde a_i, \act_{-i})  \text{ for all } \ \tilde a_i \notin \E_i \cap \F_i .    
\end{equation}
Since $\E$ is an EC, there exists a player $i,$ and an action $\hat a_i \in \E_i$ such that $\U_i(\hat a_i, \act_{-i}) > \U_i (\act),$ and 
\begin{equation} \label{eqn_EC_intersec_seven}
\U_i(\hat a_i, \act_{-i}) > \U_i ( \tilde a_i, \act_{-i}) \quad \text{for all  }\ \tilde a_i \notin  \E_i.    
\end{equation}
Now, against $\act_{-i},$ using the fact that $\E \cap \F$ satisfies the first property of an EC (see~\eqref{eqn_EC_intersec_five}), there exists another action $\bar a_i \in (\E_i \cap \F_i)$ such that 
\begin{equation} \label{eqn_EC_intersec_three}
    \U_i(\bar a_i, \act_{-i}) > \U_i ( \tilde a_i, \act_{-i}) \quad \text{for all }\ \tilde a_i \notin  \E_i \cap \F_i.
\end{equation}
Now we consider the following two sub-cases:
\begin{enumerate}[(i)]
    \item $\hat a_i \in \E_i \setminus \F_i$. Using \eqref{eqn_EC_intersec_three}, we have $\U_i(\bar a_i, \act_{-i}) > \U_i ( \hat a_i, \act_{-i}).$ Using \eqref{eqn_EC_intersec_seven} and \eqref{eqn_EC_intersec_three}, it follows that by choosing $a_i' = \bar a_i,$ \eqref{eqn_EC_intersec_six} holds.
    \item $\hat a_i \in \E_i \cap \F_i$. If $\U_i(\bar a_i, \act_{-i}) \geq \U_i ( \hat a_i, \act_{-i}),$ set $a_i' = \bar a_i,$ otherwise, set $a_i' = \hat a_i.$ It now follows from \eqref{eqn_EC_intersec_seven} and \eqref{eqn_EC_intersec_three} that this choice of $a_i'$ satisfies~\eqref{eqn_EC_intersec_six}. 
\end{enumerate}

\end{proof}


Next, we define the notion of dominant EC, inspired by the various notions of dominant pure strategy NE (see~Chapter~5 in~\cite{narahari2014game}).

\begin{defn}[Dominant equilibrium cycle]
An equilibrium cycle $\E$ is said to be dominant, if for any $a_i \notin \E_i,$ and for any $\act_{-i} \in \A_{-i},$ there exists $a_i' \in \E_i$ such that 
$$
\U_i(a_i', \act_{-i}) > \U_i ( a_i, \act_{-i}) \  \text{ and } \ \U_i(a_i', \act_{-i}) \geq 
\U_i ( \tilde a_i, \act_{-i}) \ \text{ for all } \tilde a_i \notin \E_i .
$$
\end{defn}
The following implications follow immediately.

\begin{thm} \label{thm_dominated_ECs_NEs}
Consider a game $G$.
\begin{enumerate}[(i)]
    \item If the game $G$ has a very weakly dominant pure strategy Nash equilibrium,\footnote{A strategy profile $\act^*$ is said to be a \emph{very weakly dominant Nash equilibrium} of a game~$G$ if, for any player~$i$ and for any $\act_{-i} \in \A_{-i},$  
 $\U_i(a_i^*, \act_{-i}) \geq \U_i ( \tilde a_i, \act_{-i}) \ \text{ for all } \tilde a_i \in \A_i$~\citep{narahari2014game}.} then $G$ does not have an equilibrium cycle.
    \item If $G$ has a dominant equilibrium cycle, say~$\E,$ then $\E$ is the only EC of $G$, and $G$ does not have a pure strategy Nash equilibrium.
\end{enumerate}
\end{thm}

\begin{proof}
For part (i), assume $G$ has a dominant pure strategy NE $\act^*$ as well as an EC $\E$. From Lemma~\ref{lem_no_pure_NE_in_EC}, we know that $\act^* \notin \E$, i.e., there exists a player~$i$ such that $a_i^* \notin \E_i$.  Consider the actions of opponents within the EC, i.e., $\act_{-i} \in \E_{-i}.$ Since $\act^*$ is a dominant NE, against~$\act_{-i},$ we have $\U_i(a_i^*, \act_{-i}) \geq \U_i(a_i, \act_{-i})$ for all $a_i \in \E_i,$  which contradicts the first condition of EC~$\E.$ This proves that if $G$ has a dominant pure strategy NE, then it cannot have an EC.

For part (ii), we first show that a dominant EC is unique. Suppose, for the sake of contradiction, that there exist two ECs: $\E$ and a dominant EC~$\F$. By Lemma~\ref{thm_ECs_do_not_intersect}, we have $\E \cap \F = \emptyset.$ Consider any $\act \in \E.$ Since $\F$ is a dominant EC, there exists a player~$i$, and an action $a_i' \in \F_i$ such that
$$
\U_i(a_i', \act_{-i}) \geq \U_i ( \tilde a_i, \act_{-i}) \text{ for all } \tilde a_i \notin \F_i.
$$
This contradicts the first condition of EC $\E$. This proves the uniqueness of the dominant EC. 

Next, we show that $G$ does not have any pure NE. Assume, for the sake of contradiction, that there exists a pure NE $\act^*.$ If $\act^* \in \E,$ then it would contradict Lemma~\ref{lem_no_pure_NE_in_EC}. Otherwise, if~$\act^* \notin \E,$ then there exists a player~$i$ such that $a_i^* \notin \E_i$. By definition of the dominant EC, against~$\act_{-i}^*$, there exists $a_i' \in \E_i$ such that $\U_i(a_i', \act_{-i}^*) > \U_i(a_i^*, \act_{-i}^*).$ This contradicts the assumption that $\act^*$ is an NE. Therefore, $G$ cannot have a pure strategy NE.
\end{proof}
Interestingly, the ECs identified in Propositions~\ref{lem_timing_game_EC} and~\ref{lemma:BertrandEC} for the games in Examples~\ref{ex_timing_game} and~\ref{ex_bertrand_duopoly}, respectively, are dominant (and therefore also unique, as per Theorem~\ref{thm_dominated_ECs_NEs}).

\ignore{\begin{thm}
If the game $G$ has a dominated EC, then it is unique.  
\end{thm}

\begin{proof}
With contrary, assume that there exists an EC $\E$, and a dominated EC~$\F$. From Lemma~\ref{thm_ECs_do_not_intersect}, $\E \cap \F = \emptyset.$ Consider $\act \in \E.$ Since $\F$ is a dominated EC, there exists~$i$, and  $a_i' \in \F_i$ such that $\U_i(a_i', \act_{-i}) \geq \U_i ( \tilde a_i, \act_{-i}) \text{ for all } \tilde a_i \notin \F_i.$ Using Lemma~\ref{lem_no_outside_better}, we have a contradiction to the assumption that $\E$ is an EC.
\end{proof}
}  

\ignore{
\section*{Properties}

Mostly because of the third property, which says it is the smallest interval,  we will have the following:

\begin{thm}
Each EC contains the support of at least one Nash equilibrium in mixed strategies.
\end{thm}

\section{Properties}
\begin{enumerate}
    \item We can construct an example where the union of finite closed intervals can be an EC. 
    \item In the presence of EC, there can exist NEs.
    \item 
\end{enumerate}

\begin{coro}
There exists a mixed NE with support on $\E$
\end{coro}

\begin{coro}
\begin{enumerate}[$(a)$]
    \item If payoff functions are continuous, can we prove that EC is not possible?
    \item Can we have more than one equilibrium cycle?Ye
\end{enumerate}
\end{coro}

{\color{red} Coupled game}

{\color{red}
\begin{enumerate}
    \item Continuous game, where pure NE does not exits but EC exists?
    \item Prove that Uber has dominated EC.

\end{enumerate}
}

{\color{red} Shapley polygon, and Malli Sir paper   }
} 

\section{Concluding remarks}
\label{sec:conclusion}

In classical game theory, the Nash equilibrium~\citep{nash1950equilibrium} is a foundational concept, identifying strategy profiles where no player can unilaterally deviate to improve their payoff, making it ideal to describe static one-shot interactions, and also \emph{convergent} game dynamics. Indeed, when the outcome of game dynamics is a single point (strategy profile), it is typically a (pure) Nash equilibrium; \rev{of course there are exceptions, for example, some replicator dynamics may converge to point-limits that are not Nash equilibrium (see e.g., \cite{sandholm2010population})}.   \rev{Further more, we may not even have convergence to point-limits (e.g., \cite{chain_reccurent_sets2019,benaim2012perturbations}), and this brings in the requirement for more generalized solution concepts.} For example, it might be meaningful to identify a set of points (strategy profiles) that become relevant in the long run, i.e., in the limit. In other words, we might anticipate that well-known game dynamics, such as best/better response dynamics, would converge to these sets over time, which can then be considered as the outcome of the game.

One such \emph{set-valued} outcome of game dynamics is the minimal curb set, introduced in~\cite{basu1991curb}. These sets have become essential for studying adaptive processes in games, as many adjustment processes naturally settle within a minimal curb set~\citep{minimal_curb_sets_2005, hurkens1995learning}. However, this notion assumes the existence of best responses. In this paper, we define an alternative set-valued equilibrium notion, the equilibrium cycle, that does not assume the existence of best responses, and can meaningfully be applied even in discontinuous games. The equilibrium cycle is a minimal rectangular (more formally, Cartesian product) set of action profiles that is stable against external deviations, while also being unstable with respect to internal deviations. Interestingly, minimal curb sets do not always possess the latter `unrest' property; however, \emph{non-trivial} minimal curb sets, which exclude pure Nash equilibria, do, and are equivalent to equilibrium cycles in best response games. Importantly, payoff discontinuities arise naturally in several economics applications~\citep{dasgupta1986existence, dasgupta1986existence2, reny1999existence}; these discontinuous games tend not to have best responses. In this regard, the equilibrium cycle represents a significant generalization over the curb set on non-BR games, as illustrated by several examples in this paper.

At the other extreme, specializing to finite games, we find that the equilibrium cycle is related to strongly connected sink components (or sink SCCs) of the best response graph. However, an exact equivalence does not hold here, owing primarily to the implicit rectangularity of equilibrium cycles (the same rectangularity does not hold in general for sink SCCs). It is possible to define a `non-rectangular' variation of the equilibrium cycle that is equivalent to sink SCCs of the best response graph; however, this variant is difficult to apply to discontinuous games, and also loses the equivalence to (non-trivial, minimal) curb sets in BR games. In particular, it appears that both the rectangularity as well as the closed-ness of the EC are essential for application in discontinuous games, such as those described in Examples~\ref{ex_timing_game}--\ref{ex_UBER_simialar}.

\bibliographystyle{elsarticle-harv}
\bibliography{bibliography}
\end{document}